\newif\iflink\linkfalse                 

\documentclass[12pt,twoside]{article}
\usepackage{latexsym,graphicx}
\iflink  
\def\hrefx#1#2{\href{#1}{#2} at \href{#1}{\texttt{#1}}}
\def\wkpx#1#2{\href{http://en.wikipedia.org/wiki/#1}{#2} at \href{#1}{\texttt{http://en.wikipedia/wiki/#1}}}
\def\scpx#1#2{\href{http://www.scholarpedia.org/article/#1}{#2} at \href{#1}{\texttt{http://www.scholarpedia.org/article/#1}}}
\else 
\def\hrefx#1#2{\href{#1}{#2}}
\def\wkpx#1#2{\href{http://en.wikipedia.org/wiki/#1}{#2}} 
\def\scpx#1#2{\href{http://www.scholarpedia.org/article/#1}{#2}} 
\fi

\def\paradot#1{\vspace{1.3ex plus 0.5ex minus 0.5ex}\noindent{\bf\boldmath{#1.}}}
\def\lb{\log}
\def\l{\ell}

\begin{document}

\title{\bf\Large\hrule height5pt \vskip 6mm
Algorithmic Information Theory \\[2mm]
\large [ a brief non-technical guide to the field ]
\vskip 6mm \hrule height2pt}
\author{{\bf Marcus Hutter}\\[3mm]
\normalsize RSISE$\,$@$\,$ANU and SML$\,$@$\,$NICTA \\
\normalsize Canberra, ACT, 0200, Australia \\
\normalsize \texttt{marcus@hutter1.net \ \  www.hutter1.net} }
\date{March 2007}
\maketitle

\begin{abstract}\noindent
This article is a brief guide to the field of algorithmic
information theory (AIT), its underlying philosophy, and the most
important concepts. AIT arises by mixing information theory and
computation theory to obtain an objective and absolute notion of
information in an individual object, and in so doing gives rise to
an objective and robust notion of randomness of individual objects.
This is in contrast to classical information theory that is based on
random variables and communication, and has no bearing on
information and randomness of individual objects. After a brief
overview, the major subfields, applications, history, and a map of
the field are presented.
\vspace{3ex}\def\contentsname{\centering\normalsize Contents}
{\parskip=-2.5ex\tableofcontents}
\end{abstract}

\newpage
\section{Overview}\label{secInt}

Algorithmic Information Theory (AIT) is a the
\scpx{Information\_Theory}{information theory} of individual objects,
using \scpx{Computer\_Science}{computer science}, and concerns itself
with the relationship between computation, information, and
randomness.

The information content or complexity of an object can be measured by
the length of its shortest description. For instance the string
``0101010101010101010101010101010101010101010101010101010101010101''
has the short description ``32 repetitions of '01''', while
``1100100001100001110111101110110011111010010000100101011110010110''
presumable has no simple description other than writing down the string
itself. More formally, the \scpx{Algorithmic\_Complexity}{Algorithmic ``Kolmogorov'' Complexity}
(AC) of a string $x$ is defined as the length of the shortest
program that computes or outputs $x$, where the program is run on
some fixed universal computer.

A closely related notion is the probability that a universal
computer outputs some string $x$ when fed with a program chosen
at random. This \scpx{Algorithmic\_Probability}{Algorithmic
``Solomonoff'' Probability} (AP) is key in addressing the old
philosophical problem of \scpx{Induction}{induction} in a formal way.

The major drawback of AC and AP are their incomputability.
Time-bounded ``Levin'' complexity penalizes a slow program by adding
the logarithm of its running time to its length. This leads to
computable variants of AC and AP, and \scpx{Universal\_Search}{Universal ``Levin'' Search}
(US) that solves all inversion problems in optimal (apart from some
huge multiplicative constant) time.

AC and AP also allow a formal and rigorous definition of randomness
of individual strings that does not depend on physical or
philosophical intuitions about nondeterminism or likelihood.
Roughly, a string is \scpx{Algorithmic\_Randomness}{Algorithmically
``Martin-Loef'' Random} (AR) if it is incompressible in the sense
that its algorithmic complexity is equal to its length.

AC, AP, US, and AR are the core subdisciplines of AIT, but AIT spans
into many other areas. It serves as the foundation of the
\scpx{Minimum\_Description\_Length}{Minimum Description Length} (MDL)
principle, can simplify proofs in computational complexity theory,
has been used to define a universal similarity metric between
objects, solves the Maxwell demon problem, and many others.

\section{Algorithmic ``Kolmogorov'' Complexity (AC)}\label{secAC}

\scpx{Algorithmic\_Complexity}{Algorithmic complexity}
formalizes the concept of simplicity and
complexity. Intuitively, a string is simple if it can be described
in a few words, like ``the string of one million ones'', and is
complex if there is no such short description, like for a random
string whose shortest description is specifying it bit by bit.
Typically one is only interested in descriptions or {\em codes} that
are effective in the sense that decoders are \scpx{Algorithm}{\em algorithms} on
some computer. The universal \scpx{Turing\_machine}{Turing machine} $U$ is the
standard abstract model of a general-purpose computer in theoretical
computer science. We say that program $p$ is a description of string
$x$ if $p$ run on $U$ outputs $x$, and write $U(p)=x$. The length of
the shortest description is denoted by
$$
  K(x) := \min_p\{\l(p): U(p)=x\}
$$
where $\l(p)$ is the length of $p$ measured in bits.
One can show that this definition is nearly independent
of the choice of $U$ in the sense that $K(x)$ changes by at
most an additive constant independent of $x$.
The statement and proof of this
invariance theorem in
\cite{Solomonoff:64,Kolmogorov:65,Chaitin:69} is often regarded as
the birth of algorithmic information theory.
This can be termed Kolmogorov's Thesis: the intuitive notion of
`shortest effective code' in its widest sense is captured by the
formal notion of Kolmogorov complexity, and no formal mechanism can
yield an essentially shorter code. Note that the shortest code is
one for which there is a general decompressor: the Kolmogorov
complexity establishes the ultimate limits to how short a file can
be compressed by a general purpose compressor.

There are many variants, mainly for technical reasons: The
historically first ``plain'' complexity, the now more important
``prefix'' complexity, and many others. Most of them coincide within
an additive term logarithmic in the length of the string.

\begin{figure}[tb]\normalsize
\centerline{\includegraphics[width=0.55\textwidth]{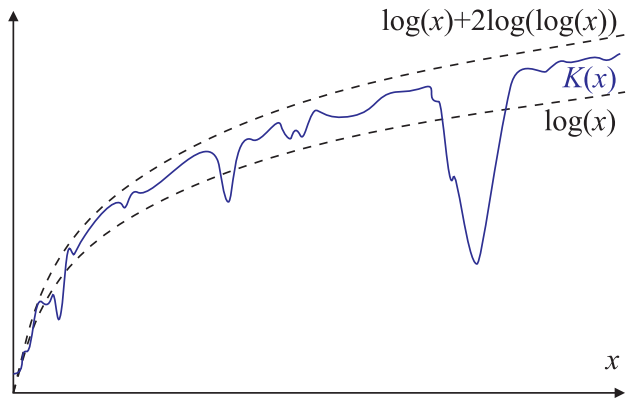}}
\vspace{1ex}
\caption{\label{figK}\it Schematic graph
of prefix Kolmogorov complexity $K(x)$ with string $x$ interpreted
as integer. $K(x)\geq\lb x$ for `most' $x$ and $K(x)\leq \lb
x+2\lb\log x+c$ for all $x$ for suitable constant $c$.}
\end{figure}

In this article we use $K$ for the prefix complexity variant. A
\scpx{Prefix\_Turing\_Machine}{prefix Turing machine} has a separate
input tape which it reads from left-to-right without backing up, a
separate worktape on which the computation takes place, and a
separate output tape on which the output is written. We define a
\scpx{Halting\_Program}{halting program} as the initial segment of
the input that is scanned at the time when the machine halts, and
the \scpx{Output}{output} is the string that has been written to the
separate output tape at that time.
The conditional prefix complexity
$$
  K(x|y):=\min_p\{\l(p):U(y, p)=x\}
$$
is the length of the shortest binary program $p\in\{0,1\}^*$ on
a universal prefix Turing machine
$U$ with output $x$ and input $y$ \cite{Li:97}.
For non-string objects (like numbers $n$, pairs of strings $(x,y)$,
or computable functions $f$) one can specify some default coding
$\langle\cdot\rangle$ and define $K(\mbox{\it
object}):=K(\langle\mbox{\it object}\rangle)$.
The most important properties are: %
\begin{itemize}\parskip=0ex\parsep=0ex\itemsep=0ex
\item that $K$ is approximable from above in the limit but not computable, %
\item the upper bounds $K(x|\l(x))\leq\l(x)$ and $K(n)\leq \lb n+2\lb\log n$, %
\item Kraft's inequality implies $\sum_x 2^{-K(x)}\leq 1$, %
\item the lower bound $K(x)\geq\l(x)$ for ``most'' $x$ %
      and $K(x)\to\infty$ for $\l(x)\to\infty$, %
\item extra information bounds $K(x|y)\leq K(x)\leq K(x,y)$, %
\item subadditivity $K(xy)\leq K(x,y)\leq K(y)+K(x|y)$, %
\item symmetry of information $K(x,y)=K(x|y,K(y))+K(y)=K(y,x)$, %
\item information non-increase $K(f(x))\leq K(x)+K(f)$ for computable functions $f$, %
\item and coding relative to a probability distribution (MDL) \\
      $K(x)\leq -\lb P(x)+K(P)$ for computable probability distributions $P$,
\end{itemize}
where all (in)equalities hold within an additive constant.
Furthermore, it shares many properties with Shannon's entropy
(information measure), but $K$ has many advantages.
The properties above allow us to draw a schematic graph of $K$ as
depicted in Figure \ref{figK}.

\section{Algorithmic ``Solomonoff'' Probability (AP)}\label{secAP}

Solomonoff (1964) considered the probability that a universal
computer outputs some string when fed with a program chosen at
random. This Algorithmic "Solomonoff" Probability (AP) is key in
addressing the old philosophical problem of induction in a formal
way. It is based on

\begin{itemize}\parskip=0ex\parsep=0ex\itemsep=0ex
\item \scpx{Occam's\_razor}{Occam's razor} (choose the simplest model consistent with the data), %
\item Epicurus' principle of multiple explanations (keep all explanations consistent with the data), %
\item Bayes's Rule (transform the a priori distribution to a posterior distribution according to the evidence, experimentally obtained data),
\item (universal) Turing machines (to compute, quantify and assign codes to all quantities of interest), and %
\item algorithmic complexity (to define what simplicity/complexity means).
\end{itemize}

Occam's razor (appropriately interpreted and in compromise with
Epicurus' principle of indifference) tells us to assign high/low a
priori plausibility to simple/complex strings $x$. Using $K$ as the
complexity measure, one could choose any monotone decreasing
function of $K$, e.g.\ $2^{-K(x)}$. The precise definition of
\scpx{Algorithmic\_Probability}{Algorithmic ``Solomonoff'' Probability}
(AP), also called universal a
priori probability, $M(x)$ is the probability that the output of a
(so-called monotone) universal Turing machine $U$ starts with $x$
when provided with fair coin flips on the input tape. Formally, $M$
can be defined as
$$
  M(x) \;:=\; \sum_{p\;:\;U(p)=x*} 2^{-\l(p)}
$$
where the sum is over all (so-called minimal, not necessarily
halting, denoted by *) programs $p$ for which $U$ outputs a string
starting with $x$. Since the shortest programs $p$ dominate the sum,
$M(x)$ is roughly $2^{-K(x)}$.

$M$ has similar remarkable properties as $K$. Additionally, the
predictive distribution
$M(x_{n+1}|x_1...x_n):=M(x_1...x_{n+1})/M(x_1...x_n)$ converges
rapidly to 1 on (hence predicts) any computable sequence
$x_1 x_2 x_3 ...$. It can also be shown that $M$ leads to excellent
predictions and decisions in general stochastic environments. If
married with sequential \scpx{Decision\_Theory}{decision theory}, it
leads to an optimal reinforcement learning agent embedded in an
arbitrary unknown environment \cite{Hutter:04uaibook}, and a formal
definition and test of intelligence.

A formally related quantity is the probability that $U$ halts when
provided with fair coin flips on the input tape (i.e.\ that a random
computer program will eventually halt). This halting probability,
also known as Chaitin's constant $\Omega$, or `the number of wisdom'
has numerous remarkable mathematical properties, and can be used for
instance to quantify Goedel's Incompleteness Theorem.

\section{Universal ``Levin'' Search (US)}\label{secUS}

Consider a problem to solve for which we have two potential
algorithms $A$ and $B$, for instance breadth versus depth first
search in a finite (game) tree. Much has been written about which
algorithm is better under which circumstances. Consider the
following alternative very simple solution to the problem: A
meta-algorithm $US$ runs $A$ and $B$ in parallel and waits for the
first algorithm to halt with the answer. Since $US$ emulates $A$ and
$B$ with half-speed, the running time of $US$ is the minimum of
$2\times$time$(A)$ and $2\times$time$(B)$, i.e.\ $US$ is as fast as
the faster of the two, apart from a factor of 2. Small factors like 2
are often minor compared to potentially much larger difference in
running time of $A$ and $B$.

\scpx{Universal\_Search}{Universal ``Levin'' Search} (US)
extends this idea from two
algorithms to {\em all} algorithms. First, since there are
infinitely many algorithms, computation time has to be assigned
non-uniformly. The optimal way is that $US$ devotes a time fraction of
$2^{-\l(p)}$ to each (prefix) program $p$. Second,
since not all programs solve the problem (some never halt, some just
print ``Hello World'', etc.) $US$ has to verify whether the output is
really a solution, and if not discard it and continue.

How does this fit into AIT?
A problem of AC $K$ is its incomputability.
Time-bounded ``Levin'' complexity penalizes a slow program by adding
the logarithm of its running time to its length:
$$
  Kt(x) \;=\; \min_p \{\l(p)+\log(\mbox{time}(p)) : U(p)=x \}
$$
It is easy to see that $Kt(x)$ is just the logarithm of the running
time (without verification) of $US$, and is therefore computable.

While universal search is nice in theory, it is not applicable in
this form due to huge hidden multiplicative constants in the running
time. Another restriction is that verification needs to be fast.
Hutter \cite{Hutter:04uaibook} developed a more general
asymptotically fastest algorithm, which removes the multiplicative
constant and necessity of verification, unfortunately at the expense
of an even larger additive constant. Schmidhuber
\cite{Schmidhuber:03newai} developed the first practical variants of
$US$ by carefully choosing the programming language ($U$),
allocating time in $US$ adaptively, designing training sequences of
increasing complexity, reusing subroutines from earlier simpler
problems, and various other ``tricks''. He also defined the Speed
Prior, which is to $Kt$ what AP is to AC.

\section{Algorithmic ``Martin-Loef'' Randomness (AR)}\label{secAR}

The mathematical formalization of the concept of probability or
chance has a long intertwined history. The (now) standard axioms of
probability, learned by all students, are due to Kolmogorov (1933).

While mathematically convincing, the semantics is far from clear.
Frequentists interpret probabilities as limits of observed relatives
frequencies, objectivists think of them as real aspects of the world,
subjectivists regard them as one's degree of belief (often elicited
from betting ratios), while Cournot only assigns meaning to events
of high probability, namely as happening for sure in our world.

None of these approaches answers the question of whether some {\em
specific individual} object or observation, like the binary strings
above, is random. Kolmogorov's axioms do not allow one to ask such
questions.

Von Mises (1919), with refinements to his approach by Wald (1937),
and Church (1940) attempted to formalize the intuitive notion of one
string looking more random than another (see the example in the
introduction) with partial success. For instance, if the relative
frequency of 1s in an infinite sequence does not converge to 1/2 it
is clearly non-random, but the reverse is not true: For instance
"0101010101..." is not random, since the pair "01" occurs too often.
Pseudo-random sequences, like the digits of $\pi$, cause
the most difficulties. Unfortunately no sequence can satisfy ''all''
randomness tests. The Mises-Wald-Church approach seemed satisfactory
untill Ville (1939) showed that some sequences are random according
to their definition and yet lack certain properties that are
universally agreed to be satisfied by random sequences. For example,
the relative frequency of `1's in increasingly long initial segments
should infinitely often switch from above 1/2 to below 1/2 and vice
versa.

Martin-Loef (1966), rather than give a definition and check whether it
satisfied all requirements, took the approach to formalize the notion
of all effectively testable requirements in the form of tests for
randomness. The tests are constructive (namely all and only lower
semi-computable) ones, which are typically all one ever cares about.
Since the tests are constructed from Turing machines, they can be
effectively enumerated according to the effective enumeration of the
Turing machines they derive from. Since the set of sequences
satisfying a test (having the randomness property the test verifies) has
measure one, and there are only countably many  tests, the set of
sequences satisfying ''all'' such tests also has measure one. These are
the ones called Algorithmic Random|Algorithmically "Martin-Loef"
Random (AR). The theory is developed for both finite strings and
infinite sequences. In the latter case the notion of test is more
complicated and we speak of sequential tests.

For infinite sequences one can show that these are exactly the
sequences which are incompressible in the sense that the algorithmic
prefix complexity of every initial segment is at least equal to
their length. More precisely, the infinite sequence
$$
  x_1 x_2 x_3... \mbox{ is AR}
  \quad\Longleftrightarrow\quad K(x_1...x_n)\geq n\;\;
  \mbox{for all suff.\ large } n
$$
an important result due to G.J. Chaitin and C. Schnorr. This notion
makes intuitive sense: A string can be compressed ''iff'' there are
some regularities in the string ''iff'' the string is non-random.

\begin{itemize}\parskip=0ex\parsep=0ex\itemsep=0ex
\item ML-random sequences cannot be effectively constructed. Yet we can
give a natural example: The \wkpx{Halting\_Probability}{halting
probability}, $\Omega$ is a real number between 0 and 1, and the
sequence of bits in its binary expansion is an infinite ML-random
sequence.
\item Randomness of other objects than strings and sequences can also be
defined.
\item Coupling the theory of AR with recursion theory (Downey and
Hirschfeldt 2007), we find a hierarchy of notions of randomness, at
least if we leave the realm of computability according to Turing. Many
variants can be obtained depending on the precise definition of
"constructive". In particular "relative randomness" based on (halting)
oracle machines leads to a rich field connected to recursion theory.
\item Finally, the crude binary separation of random versus non-random
strings can be refined, roughly by considering strings with
$K(x_1...x_n)=\alpha n$ for some $0<\alpha<1$. If strings are
interpreted as (the expansion of) real numbers, this
leads to the notion of constructive or effective Hausdorff (fractal)
dimension.
\end{itemize}

\section{Applications of AIT}\label{secAPP}

Despite the incomputability of its core concepts, AIT has many,
often unexpected, applications.

\paradot{Philosophy}
AIT helps to tackle many philosophical problems in the sense
that it allows one to formalize and quantify many intuitive but vague
concepts of great importance as we have seen above, and hence
allows one to talk about them in a meaningful and rigorous way, thus
leading to a deeper understanding than without AIT.

Most importantly, AC formalizes and quantifies the concepts of
simplicity and complexity in an essentially unique way. A core
scientific paradigm is Occam's razor, usually interpreted as ``among
two models that describe the data equally well, the simpler one
should be preferred.'' Using AC to quantify ``simple'' allowed
Solomonoff and others to develop their universal theories of
induction and action, in the field of
\scpx{Artificial\_Intelligence}{artificial intelligence}.

AIT is also useful in the foundations of thermodynamic and its
second theorem about entropy increase, and in particular for solving
the problem of \wkpx{Maxwell's\_demon}{Maxwell's demon}.

\paradot{Practice}
By (often crudely) approximating the ``ideal'' concepts, AIT has
been applied to various problems of practical interest, e.g.\ in
linguistics and genetics. The principle idea is to replace the
universal Turing machine $U$ by more limited ``Turing'' machines,
often adapted to the problem at hand. The major problem is that the
approximation accuracy is hard to assess and most theorems in AIT
break down.

The universal similarity metric by Vitanyi and others is probably
the greatest practical success of AIT: A reasonable definition for
the similarity between two objects is how difficult it is to
transform them into each other. More formally one could define the
similarity between strings $x$ and $y$ as the length of the shortest
program that computes $x$ from $y$ (which is $K(x|y)$).
Symmetrization and normalization leads to the universal similarity
metric. Finally, approximating $K$ by standard compressors like
Lempel-Ziv (zip) or bzip(2) leads to the normalized compression
distance, which has been used to fully automatically reconstruct
language and phytogenetic trees, and many other clustering
problems.

See \scpx{Applications\_of\_Algorithmic\_Information\_Theory}{Applications of
AIT} for details and references.

\paradot{Science}
In science itself, AIT can constructivize other fields:
For instance, statements in Shannon information theory and classical
probability theory necessarily only hold in expectation or with high
probability. Theorems are typically of the form ``there exists a set
of measure X for which Y holds'', i.e.\ they are useful for (large)
samples. AR on the other hand can construct high-probability
sets, and results hold for individual observations/strings.
\wkpx{Hausdorff\_Dimension}{Hausdorff dimension} and real numbers also
have constructive counterparts.

Naturally, AIT concepts have also been exploited in theoretical
computer science itself: AIT, via the incompressibility method, has
resolved many open problems in computational complexity theory and
mathematics, simplified many proofs, and is important in
understanding (dissipationless) reversible computing. It has found
applications in Statistics, Cognitive Sciences, Biology, Physics,
and Economics.

AIT can also serve as an umbrella theory for other more practical
fields, e.g., in machine learning, the Minimum Description Length
(MDL) principle can be regarded as a downscaled practical version of
AC.

\section{History, References, Notation, Nomenclature}\label{secHist}

\wkpx{Andrey\_Kolmogorov}{Andrey Kolmogorov} \cite{Kolmogorov:65}
suggested to define the information content of an object as the
length of the shortest program computing a representation of it.
\wkpx{Ray\_Solomonoff}{Ray Solomonoff} \cite{Solomonoff:64} invented
the closely related universal a priori probability distribution and
used it for time series forecasting. Together with
\wkpx{Gregory\_Chaitin}{Gregory Chaitin} \cite{Chaitin:69}, 
this initiated the field of algorithmic information theory in the
1960s. \wkpx{Leonid\_Levin}{Leonid Levin} and others significantly
contributed to the field in the 1970s (see e.g.\ \cite{Zvonkin:70}).
In particular the prefix complexity and time-bounded complexity are
(mainly) due to him.

Li and Vitanyi \cite{Li:97} is the standard AIT textbook. The book
by Calude \cite{Calude:02} focusses on AC and AR, Hutter
\cite{Hutter:04uaibook} on AP and US, and Downey and Hirschfeldt
\cite{Downey:07} on AR. The
\iflink
\else\hrefx{http://www.hutter1.net/ait.htm}{AIT website}
at http://www.hutter1.net/ait.htm
\fi
contains further references, a list of active researchers,
a mailing list, a list of AIT events, and more.

There is still no generally agreed upon notation and nomenclature in
the field. One reason is that researchers of different background
(mathematicians, logicians, and computer scientists) moved into this
field. Another is that many definitions are named after their
inventors, but if there are many inventors or one definition is a
minor variant of another, things become difficult. This article uses
descriptive naming with contributors in quotation marks.

Not even the name of the whole field is generally agreed upon.
{\it Algorithmic Information Theory}, coined by Gregory Chaitin,
seems most appropriate, since it is descriptive and impersonal, but
the field is also often referred to by the more narrow and personal
term {\it Kolmogorov complexity}.

\section{Map of the Field}\label{secMap}

The AIT field may be subdivided into about 4 separate subfields:
AC, AP, US, and AR. The fifth item below refers to applications.

\begin{itemize}
\item \scpx{Algorithmic\_Complexity}{Algorithmic ``Kolmogorov'' Complexity} (AC)
\begin{itemize}\parskip=0ex\parsep=0ex\itemsep=0ex
\item Philosophical considerations
\item Properties of AC
\item Plain (Kolmogorov) complexity
\item Prefix complexity
\item Resource bounded complexity
\item Other complexity variants
\end{itemize}

\item \scpx{Algorithmic\_Probability}{Algorithmic ``Solomonoff'' Probability} (AP)
\begin{itemize}\parskip=0ex\parsep=0ex\itemsep=0ex
\item Occam's razor and Epicurus' principle
\item Discrete algorithmic probability
\item Continuous algorithmic probability = a priori semimeasure
\item Universal sequence prediction
\item The halting probability = Chaitin's Omega = The number of Wisdom
\end{itemize}

\item \scpx{Universal\_Search}{Universal ``Levin'' Search} (US)
\begin{itemize}\parskip=0ex\parsep=0ex\itemsep=0ex
\item Levin search
\item Levin complexity and speed prior
\item Adaptive Levin search
\item Fastest algorithms for general problems
\item Optimal ordered problem solver
\item Goedel machines
\end{itemize}

\item \scpx{Algorithmic\_Randomness}{Algorithmic ``Martin-Loef'' Randomness} (AR) / Recursion Theory
\begin{itemize}\parskip=0ex\parsep=0ex\itemsep=0ex
\item Recursion theory
\item Effective real numbers
\item Randomness of reals
\item van Mises-Wald-Church randomness
\item Martin-Loef randomness
\item More randomness concepts and relative randomness
\item Effective Hausdorff Dimension
\end{itemize}

\item \scpx{Applications\_of\_AIT}{Applications of AIT}
\begin{itemize}\parskip=0ex\parsep=0ex\itemsep=0ex
\item Minimum Description/Message Length
\item Machine Learning
\item Artificial Intelligence
\item Computational Complexity
\item The Incompressibility Method
\item (Shannon) information theory
\item Reversible computing
\item Universal similarity metric
\item Thermodynamics
\item Entropy and Maxwell demon
\item Compression in nature
\end{itemize}

\end{itemize}

\paradot{Acknowledgements}
I would like to thank Paul Vit{\'a}nyi for his help
on improving the first draft of this article.


\end{document}